\documentclass[aps,prb,showpacs,showkeys,twocolumn]{revtex4}
\usepackage{bm,color,amsmath,amssymb,mathrsfs,latexsym,graphicx,psfrag}

\def\ie{{\it i.e.},\ }

\begin{document}
\title{On the linear dispersion--linear potential quantum oscillator}
\author{Martin Greiter} \affiliation{Institut f\"ur Theorie der
  Kondensierten Materie, Karlsruhe Institute of Technology, 76128
  Karlsruhe, Germany} \pagestyle{plain}

\homepage[]{http://www-tkm.physik.uni-karlsruhe.de/~greiter/}
\date{\today}

\begin{abstract}
  We solve the bi-linear quantum oscillator $H=v|p|+F|x|$ both
  quasi-classically and numerically.
\end{abstract}

\maketitle

\section{Introduction}

With the quantum theory, as it was called at the time, nearing it
first centennial anniversary, it is a rare opportunity to study an
one-dimensional ideal oscillator which has not been solved long ago.
The motion of a (non-relativistic) quantum particle with a linear
dispersion, $\epsilon_p=v\cdot |p|$, where $p=\hbar k$ is the momentum
and $v$ is a parameter, in a linearly confining potential $V(x)=F\cdot
|x|$, where $x$ is the position and the constant force $F$ again a
parameter, however, appears to provide an example.  While the problem
may look trivial at first, it is not.  The usual method of
quantization by replacing either $p\rightarrow
-i\hbar\frac{\partial}{\partial x}$ or $x\rightarrow
i\hbar\frac{\partial}{\partial p}$ cannot be applied directly, as one
cannot sensibly define the absolute value of a differential operator.

The problem is not just of academic interest, but even of relevance to
a recent experiment~\cite{lake-09np50,greiter09np5}.
Spinons, the fractionally quantized and elementary excitations in
antiferromagnetic spin chains, are well known to disperse linearly at
low energies, with $v$ proportional to the antiferromagnetic exchange
constant $J$ along the chains~\cite{Giamarchi04}. 
Spinons carry the spin of an electron but no charge.  Since the
antiparticle for a spinon is just another spinon with its spin
reversed, the spectrum has only a positive energy branch.  An one
couples two chains antiferromagnetically~\cite{Dagotto-96s618}, the
coupling $J_{\perp}$ will induce a linear confinement potential
between pairs of spinons, as the rungs between two spinons become
effectively decorrelated~\cite{Shelton-96prb8521,greiter02prb054505}.
To a very first approximation, the energy gap in the spin ladder is
hence given by the ground state energy of the bi-linear oscillator
\begin{equation}
  \label{eq:hbilin}
  H=v|p|+F|x|,
\end{equation}
which we study in this Article.  The ground state is symmetric under
one-dimensional parity $x\rightarrow -x$ and corresponds to a spinon
pair in the triplet channel, while the first excited state is
antisymmetric under $x\rightarrow -x$ corresponds to the lowest
singlet excitation in the spin ladder.  It the context of this
problem, it is hence desirable to know what the lowest eigenvalues of
\eqref{eq:hbilin} are.  From dimensional considerations, it is
immediately clear that they must scale like $\sqrt{\hbar\;\! v F}$.

\section{Quasiclassical Approach}

Even though the usual method of quantization can not be applied
directly, the problem can still be approached quasi-classically.
Applying the Bohr-Sommerfeld quantization condition~\cite{Landau3}
\begin{equation}
  \label{eq:bohr}
  \frac{1}{2\pi\hbar}\oint pdx = n+\frac{1}{2},
\end{equation}
where we are supposed to integrate over the entire classical orbit,
results with $p(x)=\frac{E_n-F|x|}{v}$ in
\begin{equation}
  \label{eq:bohr1}
  \frac{1}{2\pi\hbar}\, 4 \int_0^{E_n/F}\frac{E_n-Fx}{v}dx=n+\frac{1}{2}. 
\end{equation}
Carrying out the integration yields
\begin{equation}
  \label{eq:bohr2}
  E_n=\sqrt{\pi\left(n+\frac{1}{2}\right)}\cdot\sqrt{\hbar\;\! v F}.
\end{equation}
We expect this to constitute a reasonable approximation for the higher
energy levels, but probably not for the low lying ones.  Indeed,
this is what we will find as we solve the problem numerically below.

\section{Mathematical Formulation}

Before proceeding with the numerical solution, let us rewrite the
eigenvalue equation $H\psi(x)=E\psi(x)$ as a differential (and integral)
equation in position space.  For convenience, we consider the dimensionless
Hamiltonian 
\begin{equation}
  \label{eq:hdimless}
  H=|k|+|x|,
\end{equation}
which is obtained from \eqref{eq:hbilin} by rescaling
\begin{equation}
  \label{eq:rescale}
  \frac{H}{\sqrt{\hbar\;\!v F}}\rightarrow H,\; 
  \sqrt{\frac{\hbar\;\!v}{F}} k \rightarrow k,\; \text{and}\
  \sqrt{\frac{F}{\hbar\;\!v}} x \rightarrow x.
\end{equation}
Let us denote the eigenvalues of \eqref{eq:hdimless} by $\lambda$ and
the eigenfunctions by $\phi(x)$.  With
\begin{eqnarray}
  \label{eq:ft}
  \tilde\phi(k)&\equiv&
  \frac{1}{\sqrt{2\pi}}\int_{-\infty}^{\infty}\phi(x)  e^{-ikx}dx,\\ 
  \phi(x)&=&
  \frac{1}{\sqrt{2\pi}}\int_{-\infty}^{\infty}  \tilde\phi(k)e^{ikx}dk, 
\end{eqnarray}
we may write
\begin{eqnarray}
  \label{eq:convolution}\nonumber
  |k|\,\phi(x)
  &=&\frac{1}{\sqrt{2\pi}}\int_{-\infty}^{\infty} k\; \text{sign}(k)\,
  \tilde\phi(k)e^{ikx}dk \\[3pt]\nonumber
  &=&-i\frac{\partial}{\partial x} 
  \frac{1}{\sqrt{2\pi}}\int_{-\infty}^{\infty}\text{sign}(k)\,
  \tilde\phi(k)e^{ikx}dk \\[3pt] 
  &=& -i\frac{\partial}{\partial x}
  \frac{1}{\sqrt{2\pi}}\int_{-\infty}^{\infty} \tilde s(x-x')\,\phi(x') dx',
\end{eqnarray}
where
\begin{equation}
  \label{eq:sgn}\nonumber
  \text{sign}(k)=
  \begin{cases}
    +1  &k\ge 0\\
    -1 &k<0
  \end{cases}
\end{equation}
is the sign function and 
\begin{eqnarray}
  \label{eq:stilde}\nonumber
  \tilde s(x)
  &=&\frac{1}{\sqrt{2\pi}}\lim_{\epsilon\rightarrow 0}\int_{-\infty}^{\infty}
  \text{sign}(k)\,e^{-\epsilon |k|}e^{ikx}dk \\[3pt] 
  &=&\frac{2i}{\sqrt{2\pi}}\lim_{\epsilon\rightarrow 0}\frac{x}{x^2+\epsilon^2}
  \,=\frac{2i}{\sqrt{2\pi}}\,\mathcal{P}\frac{1}{x},
\end{eqnarray}
where $\mathcal{P}$ denotes the principal part, is the Fourier
transform thereof.  The eigenfunctions $\phi(x)$ with eigenvalues
$\lambda$ of \eqref{eq:hdimless} are hence the solutions of
\begin{equation} 
  \label{eq:diffeqn}
  \frac{1}{\pi}\frac{\partial}{\partial x}
  \mathcal{P}\!\int_{-\infty}^{\infty}\frac{\phi(x')}{x-x'}dx'
  +|x|\,\phi(x)=\lambda\phi(x).
\end{equation} 
While \eqref{eq:diffeqn} provides a clear mathematical formulation of
the problem, we are not aware of any method to solve it analytically,
nor consider it a viable starting point for numerical work.

\section{Numerical Solution}

To solve \eqref{eq:hdimless} numerically, we exactly diagonalize
a finite Hamiltonian matrix we obtain through discretization of position
space with a suitably chosen cutoff.


Let this discrete Hilbert space consist of $N$ sites, with the
positions 
\begin{equation}
  \label{eq:xi}
   x_i=a\left(i-\frac{N+1}{2}\right)
\end{equation}
where $i=1,2,\ldots N$ and 
$a$ is the lattice constant.
The cutoff $|x_c|={Na}/{2}$ 
in real space implies a cutoff 
\begin{equation}
  \label{eq:lambdac}
  \lambda_c=\frac{Na}{2} 
\end{equation}
for the potential energy in \eqref{eq:hdimless}, which must be chosen
significantly larger than the largest eigenvalue $\lambda_n$ we wish
to evaluate reliably.  (From \eqref{eq:bohr2}, we expect $\lambda_n$
to be of order $\sqrt{\pi\left(n+\frac{1}{2}\right)}$.)  On the other
hand, the classically allowed part of the Hilbert space will contain
only of the order of ${N}/{\lambda_c}$
sites for the ground state, which implies that we must further 
require $\lambda_c\ll N$.

The lattice provides us simultaneously with a cutoff in momentum
space, $-\pi\le ak\le\pi$.  We may hence expand $|k|$ in a Fourier
series,
\begin{equation}
  \label{eq:fourier}
  |ak|=\frac{b_0}{2}+\sum_{m=1}^{\infty} b_m \cos(mak)
\end{equation}
with
\begin{equation}
  \label{eq:cm}
  b_m=\frac{1}{\pi}\int_{-\pi}^{\pi}dk |k| \cos(mk)=
  \begin{cases}
    \pi                         &m=0\\
    -\frac{4}{\pi}\frac{1}{m^2} &m\ \text{odd}\\
    0                           &\text{otherwise},
  \end{cases}
\end{equation}
as one may easily verify through integration by parts.  We
proceed by writing \eqref{eq:hdimless} in second quantized notation,
\begin{eqnarray}
  \label{eq:h2}\nonumber
  H &=& \sum_k |k|\, c_k^{\dagger}c_k + \sum_i |x_i|\, c_i^{\dagger}c_i\\ 
     &=& \frac{1}{a} \sum_k |ak|\, c_k^{\dagger}c_k 
     + a \sum_i \left|i-\frac{N+1}{2}\right|\, c_i^{\dagger}c_i\quad 
\end{eqnarray}
where 
\begin{equation}
  \label{eq:ftop}
  c_k^{\dagger}=\frac{1}{\sqrt{N}}\sum_i e^{ikx_i} c_i^{\dagger},\quad
  c_i^{\dagger}=\frac{1}{\sqrt{N}}\sum_k e^{-ikx_i} c_k^{\dagger}.
\end{equation}
Since 
\begin{equation}
  \label{eq:cos}
  \sum_k\cos(mak)\, c_k^{\dagger}c_k
  = \frac{1}{2}\sum_i (c_i^{\dagger}c_{i+m} + \text{h.c.}),
\end{equation}
we obtain
\begin{equation}
  \label{eq:hfinal}
  H=\sum_{i,j=1}^{N}c_i^{\dagger}h_{ij}c_j
\end{equation}
with 
\begin{equation}
  \label{eq:hij}
  h_{ij}=  \begin{cases}
    \frac{N}{2\lambda_c}\frac{\pi}{2} 
    + \frac{2\lambda_c}{N}\left|i-\frac{N+1}{2}\right|&i=j\\[3pt]
    - \frac{N}{2\lambda_c}\frac{2}{\pi}\frac{1}{(i-j)^2}&i-j\ \text{odd}\\[3pt]
    0                           &\text{otherwise},
  \end{cases}
\end{equation}
where we have substituted $\frac{2\lambda_c}{N}$ for $a$.

\begin{table}[t]
  \centering
  \begin{tabular}{c@{\hspace{20pt}}c@{\hspace{20pt}}c@{\hspace{20pt}}c@{\hspace{20pt}}c}
    \hline\hline
    $m$ & $\lambda_{2m}$ & $\lambda_{2m+1}$ & $\lambda_{2m}$ & $\lambda_{2m+1}$ \\
&\multicolumn{2}{c@{\hspace{20pt}}}{numerically}
&\multicolumn{2}{c}{quasi-classically}\\ 
   \hline
   0  &  1.10408 &  2.23229 &  1.2533 &  2.1708\\  
   1  &  2.77281 &  3.33002 &  2.8025 &  3.3160\\ 
   2  &  3.75118 &  4.16416 &  3.7599 &  4.1568\\ 
   3  &  4.51300 &  4.85855 &  4.5189 &  4.8541\\ 
   4  &  5.16402 &  5.46623 &  5.1675 &  5.4631\\ 
   5  &  5.74065 &  6.01303 &  5.7434 &  6.0107\\ 
   6  &  6.26457 &  6.51426 &  6.2666 &  6.5124\\ 
   7  &  6.74763 &  6.97965 &  6.7493 &  6.9782\\ 
   8  &  7.19841 &  7.41595 &  7.1997 &  7.4147\\ 
   9  &  7.62246 &  7.82800 &  7.6236 &  7.8269\\
  \hline\hline
  \end{tabular}
  \caption{Eigenvalues $\lambda_n$ for $n=0,\ldots,19$ obtained by exact
    diagonalization of \eqref{eq:hij} for $N=20001$, $\lambda_c=20$.
    From the scaling behavior with $N$ and comparisons of different values 
    for $\lambda_c$, we estimate the error due to the finite size to be 
    less than $\pm 0.00002$ for $n$ even and $\pm 0.00001$ for $n$ odd.  
    For comparison, we also list the quasi-classical values 
    \eqref{eq:bohr2}.}
    \label{tab:lambda}
\end{table}

\begin{figure}[t]
  \includegraphics[width=\linewidth]{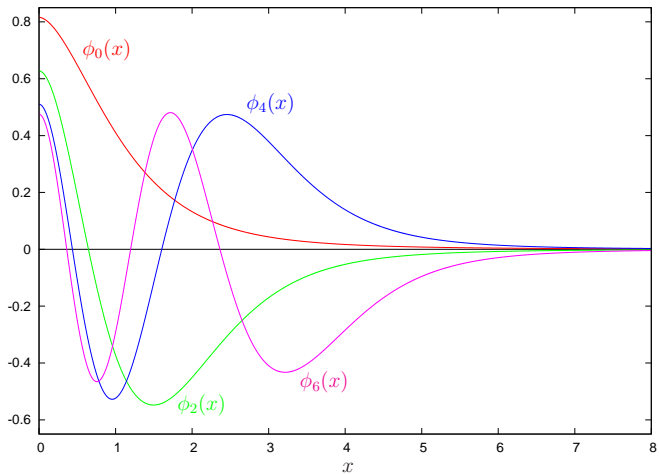}
  \caption{(Color online) The first four symmetric eigenfunctions
    $\phi_n(-x)=\phi_n(x)$ for $n$ even obtained numerically for $N=20001$,
    $\lambda_c=20$.}
  \label{fig:psi-even}
\end{figure}

\begin{figure}[t]
  \includegraphics[width=\linewidth]{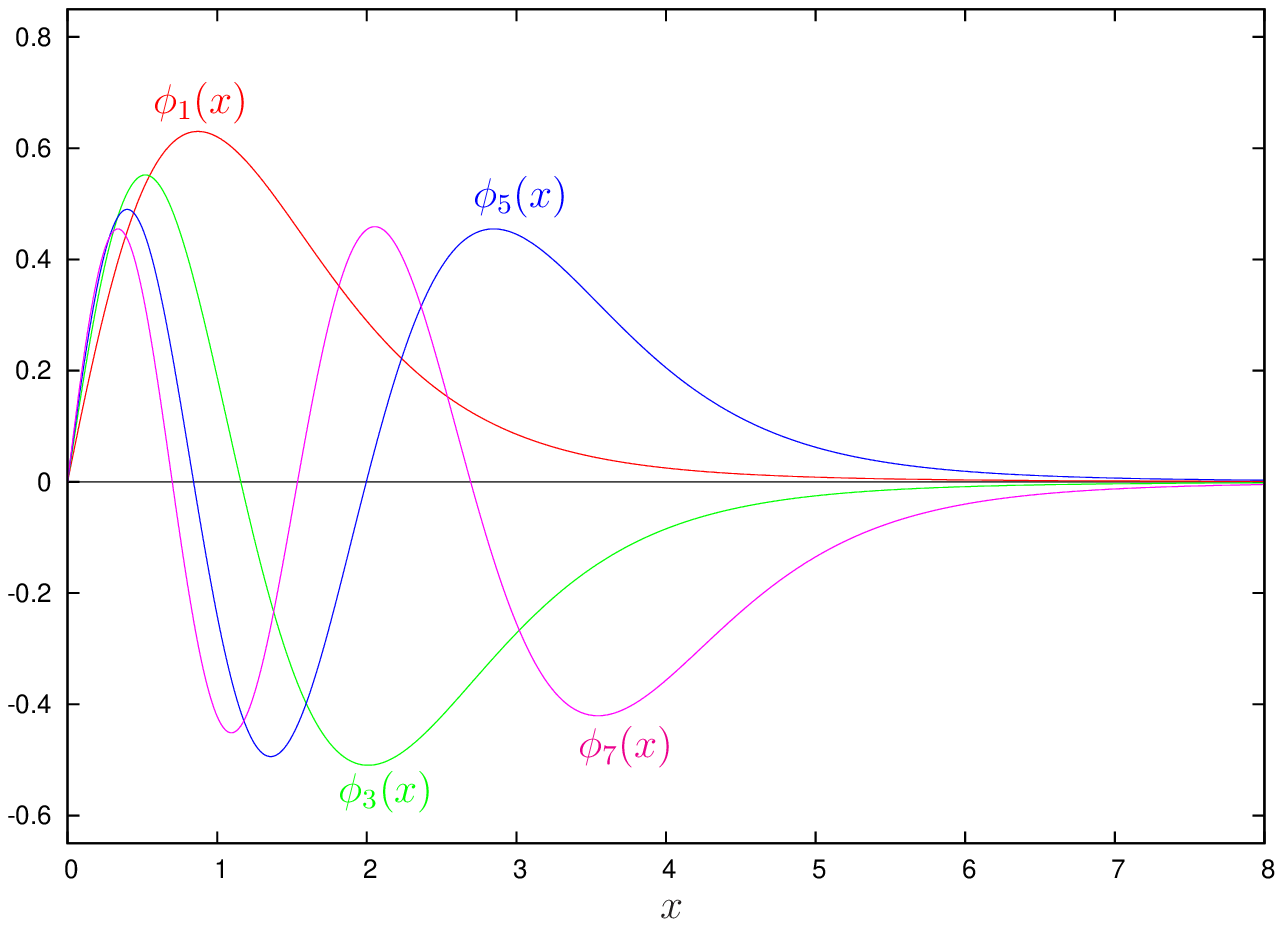}
  \caption{(Color online) The first four antisymmetric eigenfunctions
    $\phi_n(-x)=-\phi_n(x)$ for $n$ odd obtained numerically for $N=20001$,
    $\lambda_c=20$.}
  \label{fig:psi-odd}
\end{figure}

Numerical diagonalization of $h_{ij}$ yields the eigenvalues
$\lambda_n$ and eigenfunctions $\phi_n(x_i)$ of \eqref{eq:hdimless}, and
hence the eigenvalues 
and eigenfunctions 
\begin{equation}
  \label{eq:ebilin}
  E_n=\lambda_n \sqrt{\hbar\;\! v F},\ \ 
  \psi_n(x)=\phi_n\!\!\left(\sqrt{\frac{F}{\hbar\;\!v}}\, x\right)
\end{equation}
of \eqref{eq:hbilin}.
The results for $N=20001$, $\lambda_c=20$ are listed in Table
\ref{tab:lambda} and Figures \ref{fig:psi-even} and \ref{fig:psi-odd}.
(We have chosen an odd number for $N$, because this means that the
position $x=0$, where the potential $|x|$ is not differentiable,
coincides with a lattice point.  Including this point improves the
convergence of the eigenvalues and functions for $n$ even.)  From
Table \ref{tab:lambda}, we see that the quasi-classically obtained
eigenvalues converge towards the numerically obtained values as 
$n$ is increased.

\begin{figure}[t]
  \includegraphics[width=\linewidth]{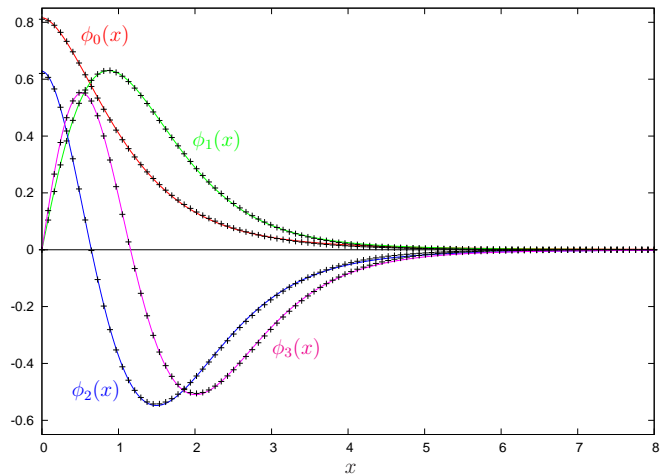}
  \caption{(Color online) Juxtapositions of the first four eigenfunctions
    $\phi_n(x)$ obtained numerically (lines) with the fits described 
    in the text (black crosses).}
  \label{fig:psi-approx}
\end{figure}

\begin{table}[t]
  \centering
  \begin{tabular}{c@{\hspace{20pt}}c@{\hspace{20pt}}c@{\hspace{20pt}}c@{\hspace{20pt}}c}
    \hline\hline
    $n$ & $a_n$ & $b_n$ & $c_n$ & $d_n$ \\
   \hline
   0  & 1.1849 & 0.57196 & 0.4681 &   \\  
   1  & 1.7443 & 0.96843 & 1.9494 &   \\ 
   2  & 1.9517 & 0.94194 & 2.2398 & 0.64431  \\ 
   3  & 2.2842 & 1.17617 & 2.9428 & 1.15453  \\ 
  \hline\hline
  \end{tabular}
  \caption{Parameters obtained numerically from fitting \eqref{eq:phi01fit}
    and \eqref{eq:phi23fit} to the functions $\phi_n(x)$ obtained by
    exact diagonalization of \eqref{eq:hij} for $N=20001$, $\lambda_c=20$.}
    \label{tab:fits}
\end{table}

The eigenfunctions obtained numerically can be approximated by
\begin{equation}
  \label{eq:phi01fit}
  \phi_n(x)=x^n\exp\Bigl(-a_n\sqrt{x^2+b_n^2}+c_n\Bigr)
\end{equation}
for $n=0,1$ and by
\begin{equation}
  \label{eq:phi23fit}
  \phi_n(x)
  =x^{n-2}\left(d_n^2-x^2\right)\exp\Bigl(-a_n\sqrt{x^2+b_n^2}+c_n\Bigr)
\end{equation}
for $n=2,3$, with parameters $a_n$, $b_n$, $c_n$, and $d_n$ listed in
Table \ref{tab:fits}.  Comparisons of these fits to the numerically
obtained eigenfunctions are shown in Figure \ref{fig:psi-approx}.  The
fits are not as good an approximation as Figure \ref{fig:psi-approx}
may suggest, however, as they fall off as $\exp(-a|x|)$ while the true
eigenfunctions $\phi_n(x)$ fall off as $1/x^3$ for $n$ even and as
$1/x^4$ for $n$ odd as $x\rightarrow\infty$.

This asymptotic behavior of the eigenfunctions can be understood
physically through second order perturbation theory.  If we consider a
small region around a point $x\gg \lambda$ (\ie very far away from the
classically allowed region for the eigenstate with energy $\lambda$),
the amplitude there will be governed by scattering into this region
from the classically allowed region, which contains almost the entire
amplitude of the state.  From \eqref{eq:hij}, this scattering is
proportional to 
\begin{equation}
  \label{eq:scatt}
  \int_{-\lambda_n-\lambda_t}^{\lambda_n+\lambda_t}\frac{\phi_n(x')}{(x-x')^2} dx' 
  \propto
  \begin{cases}
    \frac{1}{x^2}  &n\ \text{even}\\[3pt]
    \frac{1}{x^3}  &n\ \text{odd},
  \end{cases}
\end{equation}
where $\lambda_t$ is a cutoff to insure that we include the tail
immediately surrounding the classically allowed region in the integral
(from Figs.\ \ref{fig:psi-even} and \ref{fig:psi-odd}, we see that
$\lambda_t=3$ would be a reasonable choice).  With the potential
energy in the region we consider given by $|x|$, the amplitude for
finding the particle there will be proportional to $1/x^3$ for $n$
even and as $1/x^4$ for $n$ odd.

The numerical work reported here indicates that, within the limits
of accuracy, the solutions are differentiable at $x=0$, \ie the
expansion of $\phi_n(x)$ around $x=0$ does not contain a term
proportional to $|x|$ for $n$ even or $x\,|x|$ for $n$ odd.
Unfortunately, we have not been able to reach a conclusion regarding
higher terms, and cannot tell whether there are terms proportional to
$x^2|x|$ for $n$ even or $x^3|x|$
for $n$ odd.

\section{Further Considerations}

It would be highly desirable to identify the exact eigenvalues and
functions of \eqref{eq:hdimless}.  Unfortunately, we have as of yet
not even succeeded in obtaining those for the ground state.  A few
thoughts on this problem, however, are possibly worth mentioning.

\subsection{Fourier Symmetry}
As the Hamiltonian \eqref{eq:hdimless} maps onto itself under Fourier
transformation, and all the eigenstates are non-degenerate, the
eigenfunctions $\phi(x)$ must likewise map into itself under Fourier
transformation \eqref{eq:ft},
\begin{equation}
  \label{eq:ftequiv}
  \tilde\phi_n(x) = (-i)^n\phi_n(x).
\end{equation}
This condition is directly fulfilled by certain functions, like the
Gaussian eigenfunctions of the harmonic oscillator $H=\frac{1}{2}(k^2+x^2)$,
\begin{equation}\nonumber
  \label{eq:harm}
  \phi_n(x)
  =\left(x-\frac{\partial}{\partial x}\right)^n\exp\left(-\frac{x^2}{2}\right),
\end{equation}
or the function
\begin{equation}\nonumber
  \phi_0(x)=\frac{1}{\cosh\left(\sqrt{\frac{\pi}{2}}x\right)}.\label{eq:sech}
\end{equation}
The eigenfunctions of
\eqref{eq:hdimless}, however, do not need to be of any such particular
form.  For example, the Ansatz
\begin{equation}
  \label{eq:ftplus}
  \phi_n(x) =i^n\tilde\varphi_n(x)+\varphi_n(x)
\end{equation}
satisfies \eqref{eq:ftequiv} in general, as \eqref{eq:ft} implies
${\tilde{\tilde\varphi}}_n(x)=\varphi_n(-x)=(-1)^n\varphi_n(x)$.
%
%

It is conceivable that the function $\varphi(x)$ displays the required
asymptotic behavior mentioned above,
while the Fourier transform $\tilde\varphi(x)$ falls off more
rapidly.  A first guess for the ground state along these lines might
be
\begin{equation}
  \label{eq:phi1}
  \varphi_0(x)=\frac{1}{(x^2+a^2)^{3/2}},
\end{equation}
with its Fourier transform given by a modified Bessel function of the
second kind,
\begin{equation}
  \label{eq:phi1}
  \tilde\varphi_0(x)=\sqrt{\frac{2}{\pi}}\frac{|x|}{a} K_1(a |x|).
\end{equation}
With $a\approx 1.172$, this provides a very reasonable approximation,
but does not solve the problem exactly.

\subsection{Asymptotic Behavior}
    
Even though we are unable to solve \eqref{eq:diffeqn}, we can use it
to determine the asymptotic behavior of the solutions $\phi_n(x)$ as
$x\rightarrow\infty$ accurately.  Let us first consider even
eigenfunctions $\phi_n(-x)=\phi_n(x)$.  Then \eqref{eq:diffeqn}
becomes
\begin{equation} 
  \label{eq:diffeqneven1}
  \frac{1}{\pi} \frac{\partial}{\partial x}
  \mathcal{P}\!\int_{0}^{\infty}\frac{2x\phi_n(x')}{x^2-x'^2}dx'
  +|x|\,\phi_n(x)=\lambda_n\phi_n(x),
\end{equation} 
For $x\rightarrow+\infty$, we obtain
\begin{equation} 
  \label{eq:diffeqneven2}
  -\frac{2}{\pi}\frac{1}{x^2} \int_{0}^{\infty}\phi_n(x')dx'
  +\,\text{O}\Bigl(\frac{1}{x^4}\Bigr)
  +(x-\lambda_n)\,\phi_n(x)=0.
\end{equation} 
With \eqref{eq:ft} and \eqref{eq:ftequiv}, however, we may write
\begin{equation}
  \label{eq:inteven}
  \int_{-\infty}^{\infty}\phi_n(x)dx
  =\sqrt{2\pi}\tilde\phi_n(0)
  =(-i)^n\sqrt{2\pi}\phi_n(0),
\end{equation}
and hence obtain for $n$ even
\begin{equation} 
  \label{eq:diffeqneven3}
  \phi_n(x)= 
  (-1)^{n/2}
  \sqrt{\frac{2}{\pi}}\phi_n(0) 
  \biggl(\frac{1}{x^3}+\frac{\lambda_n}{x^4}+
  \,\text{O}\Bigl(\frac{1}{x^5}\Bigr)\!\biggr).
\end{equation}

\vspace{10pt}
Similarly, we write \eqref{eq:diffeqn} for the odd eigenfunctions
$\phi_n(-x)=-\phi_n(x)$
\begin{equation} 
  \label{eq:diffeqnodd1}
  \frac{1}{\pi} \frac{\partial}{\partial x}
  \mathcal{P}\!\int_{0}^{\infty}\frac{2x'\phi_n(x')}{x^2-x'^2}dx'
  +|x|\,\phi_n(x)=\lambda_n\phi_n(x),
\end{equation} 
For $x\rightarrow+\infty$, we obtain
\begin{equation} 
  \label{eq:diffeqnodd2}
  -\frac{4}{\pi}\frac{1}{x^3} \int_{0}^{\infty}x'\phi_n(x')dx'
  +\text{O}\Bigl(\frac{1}{x^5}\Bigr)
  +(x-\lambda_n)\,\phi_n(x)=0
\end{equation} 
With \eqref{eq:ft} and \eqref{eq:ftequiv}, the integral becomes
\begin{eqnarray}
  \label{eq:intodd}\nonumber
  \int_{-\infty}^{\infty}x\,\phi_n(x)dx
  &=&\sqrt{2\pi}\cdot
  i\frac{\partial}{\partial k}\tilde\phi_n(k)\Bigl|_{k=0}\Bigr.\\
  &=&(-i)^{n-1}\sqrt{2\pi}\phi_n'(0).
\end{eqnarray}
This yields for $n$ odd
\begin{equation} 
  \label{eq:diffeqnodd3}
  \phi_n(x)= 
  (-1)^{\frac{\scriptstyle (n-1)}{\scriptstyle 2}}
  2\sqrt{\frac{2}{\pi}}\phi_n'(0) 
  \biggl(\frac{1}{x^4}+\frac{\lambda_n}{x^5}+
  \,\text{O}\Bigl(\frac{1}{x^6}\Bigr)\!\biggr).
\end{equation} 
The asymptotic behavior emphasizes how different the bi-linear
oscillator \eqref{eq:hdimless} is from the well known harmonic
oscillator.

\subsection{Integral relations}

We can apply some general properties of Hilbert transformations,
defined as\cite{Erdeyli54}
\begin{equation}
  \label{eq:hilbert}
  \mathcal{H}[f](x)\equiv\frac{1}{\pi} 
  \mathcal{P}\!\int_{-\infty}^{\infty}\frac{f(x')}{x-x'}dx',
\end{equation}
where $\mathcal{P}$ denotes the principle part, to rewrite
\eqref{eq:diffeqn}.  With
\begin{eqnarray}
  \label{eq:hilbertp}
  \frac{\partial}{\partial x}\mathcal{H}[f](x)&=&\mathcal{H}[f'](x),
  \\[5pt]
  \mathcal{H}\big[\mathcal{H}[f]\big](x)&=&-f(x),
\end{eqnarray}
we obtain
\begin{equation}
  \label{eq:hilbertdiffeqn}
  \frac{\partial\phi_n(x)}{\partial x}+ \frac{1}{\pi} 
  \mathcal{P}\!\int_{-\infty}^{\infty}\frac{(\lambda_n-|x'|)\phi_n(x')}{x-x'}dx'
  =0.
\end{equation}

Expanding the integral for the limit $x\rightarrow\infty$, we obtain
for $n$ even
\begin{equation}
  \label{eq:hilberteven1}
  \frac{\partial\phi_n(x)}{\partial x}= \frac{2}{\pi}\frac{1}{x} 
  \int_{0}^{\infty}(\lambda_n-x')\phi_n(x')dx'
  +\,\text{O}\Bigl(\frac{1}{x^3}\Bigr).
\end{equation}
With \eqref{eq:diffeqneven3}, this implies 
\begin{equation}
  \label{eq:hilberteven2}
  \int_{0}^{\infty}(x-\lambda_n)\phi_n(x)dx=0,
\end{equation}
and with \eqref{eq:inteven}
\begin{equation}
  \label{eq:hilberteven3}
  \int_{0}^{\infty}x\phi_n(x)dx=(-1)^{n/2}
  \sqrt{\frac{\pi}{2}}\lambda_n\phi_n(0).
\end{equation}

Similarly, we obtain in this limit for $n$ odd
\begin{equation}
  \label{eq:hilbertodd1}
  \frac{\partial\phi_n(x)}{\partial x}=\frac{2}{\pi}\frac{1}{x^2} 
  \int_{0}^{\infty}x'(\lambda_n-x')\phi_n(x')dx'
  +\,\text{O}\Bigl(\frac{1}{x^4}\Bigr).
\end{equation}
With \eqref{eq:diffeqnodd3}, this implies 
\begin{equation}
  \label{eq:hilbertodd2}
  \int_{0}^{\infty}x(x-\lambda_n)\phi_n(x)dx=0,
\end{equation}
and with \eqref{eq:intodd}
\begin{equation}
  \label{eq:hilbertodd3}
  \int_{0}^{\infty}x^2\phi_n(x)dx
  =(-1)^{\frac{\scriptstyle (n-1)}{\scriptstyle 2}}
  \sqrt{\frac{\pi}{2}}\lambda_n\phi_n(0).
\end{equation}
\\

\section{Conclusion}

We have succeeded in solving the bi-linear oscillator $H=v|p|+F|x|$
both quasi-classically and numerically.  In an attempt to solve it
analytically as well, we have derived a differential and integral
equation, and obtained the asymptotic behavior for large $x$.  We
further formulated several conditions the solutions must satisfy.  The
problem of obtaining an analytical solution, however, is still open.

%

\begin{acknowledgments}
  I am grateful to R.\ von Baltz, W.\ Lang, A.D.\ Mirlin, and P.\ W\"olfle 
  for valuable discussions of this problem.
\end{acknowledgments}


\end{document}